\documentclass[twocolumn,numberedappendix,twoside]{emulateapj}

\usepackage{lscape}
\usepackage{natbib}
\usepackage{rotating}

\def\ie{i.e.}
\def\eg{e.g.}
\def\deg{\ifmmode^\circ\else$^\circ$\fi}
\def\Msun{\ifmmode{\mathcal{M}_\odot}\else{$\mathcal{M}_\odot$}\fi}
\def\Lsun{\ifmmode{\mathrm{L}_\odot}\else{L$_\odot$}\fi}
\def\Massstar{\ifmmode{\mathcal{M}_{\ast}}\else{$\mathcal{M}_\ast$}\fi}

\def\zf{\ifmmode{z_{\rm f}}\else{$z_{\rm f}$}\fi}
\def\Ks{\ifmmode{K_{\rm s}}\else{$K_{\rm s}$}\fi}
\def\Mcut{\ifmmode{M_{\rm cut}}\else{$M_{\rm cut}$}\fi}
\def\ncmod{{\tt NCMOD\/}}

\defcitealias{2010arXiv1002.3537E}{Paper I}
\defcitealias{2009A&A...501..505L}{LSJ09a}
\defcitealias{2009ApJ...694..643L}{LSJ09c}
\slugcomment{revised manuscript, \today}

\slugcomment{02 March 2010 version}

\shortauthors{Eliche-Moral et al.}
\shorttitle{The role of wet, mixed, and dry major mergers in the buildup of mETGs at $z\sim 1$}
\begin{document}

\title{On the buildup of massive early-type galaxies at $z\lesssim 1$. \\
II- The coordinated key role of wet, mixed, and dry major mergers}

\author{M.~Carmen Eliche-Moral\altaffilmark{1}, Mercedes Prieto\altaffilmark{2,3}, Jes\'{u}s Gallego\altaffilmark{1}, and Jaime Zamorano\altaffilmark{1}}

\altaffiltext{1}{Departamento de Astrof\'{\i}sica, Universidad Complutense de Madrid, 28040 Madrid, Spain} 
\altaffiltext{2}{Instituto de Astrof\'{\i}sica de Canarias, C/ V\'{\i}a L\'actea, E-38200 La Laguna, Canary Islands, Spain} 
\altaffiltext{3}{Departamento de Astrof\'{\i}sica, Universidad de La Laguna, 
Avda.\ Astrof\'{\i}sico Fco.\ S\'anchez, E-38200 La Laguna, Canary Islands, 
Spain\\
{\tt e-mail\/}: mceliche@fis.ucm.es}

\begin{abstract}
Hierarchical models predict that present-day massive early-type galaxies (mETGs) have finished their assembly at a quite late cosmic epoch ($z\sim 0.5$), conflicting directly with galaxy mass-downsizing. In \citet{2010arXiv1002.3537E}, we presented a semi-analytical model that predicts the increase by a factor of $\sim 2.5$ observed in the number density of mETGs since $z\sim 1$ to the present, just accounting for the effects of the major mergers strictly-reported by observations. Here, we describe the relative, coordinated role of wet, mixed, and dry major mergers in driving this assembly. Accordingly to observations, the model predicts that: 1) wet major mergers have controlled the mETGs buildup since $z\sim 1$, although dry and mixed mergers have also contributed significantly to it; 2) the bulk of this assembly takes place during the $\sim 1.4$\,Gyr time-period elapsed at $0.7<z<1$, being nearly frozen at $z\lesssim 0.7$; 3) this frostbite can be explained just accounting for the observational decrease of the major merger fraction since $z\sim 0.7$, implying that major mergers (and, in particular, dry events) have contributed negligibly to the mETGs assembly during the last $\sim 6.3$\,Gyr; and 4) major mergers are responsible for doubling the stellar mass at the massive-end of the red sequence since $z\sim 1$, accounting dry mergers for $\sim 35$\% of the mass accretion of local mETGs in the last $\sim 8$\,Gyr. The model can predict the fractions of present-day E's and S0-S0a's if the last ones derive exclusively from a wet major merger occurred since $z\sim 1$, while E's have also undergone at least one gas-poor event during the same epoch. It can also reproduce the local percentages of rapid-rotating, disky and slow-rotating, boxy E's if the last major merger assembling the former ones is mixed and that of the last ones is dry. The most striking model prediction is that at least $\sim 87(^{110}_{67})$\% of the mETGs existing at $z\sim 1$ are not the passively-evolved, high-z counterparts of present-day mETGs, but their gas-poor progenitors instead. This implies that $\gtrsim 95$\% of local mETGs have acquired $\gtrsim 1/3$ of their present stellar masses through major mergers during the last $\sim 8$\,Gyr, meaning that $\lesssim 5$\% of present-day mETGs have been really in place since $z\sim 1$. So, the model derives a redshift of final assembly for present-day mETGs in agreement with hierarchical models ($z\sim 0.5$), reproducing at the same time the observed buildup of mETGs at $z\lesssim 1$. The model proves that the late epoch of final mETGs buildup predicted by hierarchical models and the galaxy mass-downsizing phenomenon can be reconciled, just through a correct interpretation of observational results.
\end{abstract}


\section{Introduction}
\label{sec:introduction}

Massive galaxies seem to have been in place in the Universe before than their low-mass counterparts. This observational phenomenon, known as galaxy mass-downsizing, indicates that galaxies with $\mathcal{M}_{*} \gtrsim 10^{11}\Msun$ must have finished their assembly at $z\sim 0.8$ \citep{2008ApJ...675..234P}. This epoch coincides with the moment at which early-type galaxies (E-S0a's, ETGs hereafter) start to dominate the massive end of the galaxy luminosity function \citep[LFs, see][]{2004ApJ...608..752B}, suggesting "the existence of a dominant mechanism that links the shutdown of star formation (SF) and the acquisition of a spheroidal morphology in massive systems" \citep{2010ApJ...709..644I}. 

Present hierarchical models of galaxy formation derive that this mechanism has been major merging. $\Lambda$CDM models predict that massive ETGs with local masses $\mathcal{M}_{*} \gtrsim 10^{11}\Msun$ (mETGs) must be the final remnants of the most "rich and violent" merging trees occurred in the Universe, being the latest systems to be completely in place into the cosmic scenario \citep[at $z\sim 0.5$, see][]{1999AJ....117.1651G,2002ApJ...568...52W,2003ApJ...590..619M,2005Natur.435..629S,2006MNRAS.366..499D,2006ApJ...636L..81N,2009A&A...497...35G,2009ApJ...702..307S}. Obviously, this late epoch for the complete assembly of mETGs conflicts directly with observational mass-downsizing. However, the merger-origin of mETGs is supported by N-body simulations in the last decade, which show that major mergers between galaxy with typical gas-contents (of $\lesssim 20$-30\% of their baryonic masses) give place to E-S0a remnants basically \citep[][]{1996ApJ...471..115B,2004A&A...418L..27B,2005A&A...437...69B,2003ApJ...597..893N,2006ApJ...636L..81N}. 

The major merger-origin of ETGs is also suggested by numerous observational results. The properties of the nuclei of disk-disk merger remnants are coherent with their transformation into normal E-S0's after 1-2\,Gyrs, the majority of local mETGs and of the brightest group galaxies at intermediate redshifts exhibit traces of having experienced a major merger, and the structure and concentration of the most massive ETGs suggest a formation through radial mergers \citep[see][]{1995A&A...298..405S,1997ApJ...486L..87B,2003MNRAS.341..747B,2003ApJ...597..893N,2005ASSL..329..143S,2005AJ....130.2647V,2006ApJ...636L..81N,2006ApJ...641...90R,2007MNRAS.379..401E,2007AJ....134.2124R,2008ApJ...684.1062F,2008ApJ...683L..17T,2008ApJ...688...48V,2008A&A...491..713W,2009ApJ...691L.138S,2010MNRAS.tmp...24B}. Moreover, observational studies have identified systems that could correspond to intermediate stages of major mergers evolving into ETGs \citep[such as the K$+$A galaxies, the blue ETGs, the red spirals, or the post-starburst galaxies, see][]{2004AJ....128.1990C,2005ApJ...635..243F,2007MNRAS.382..960K,2010arXiv1002.3076H,2009MNRAS.397.1940F,2009MNRAS.398.1651H,2009AJ....138..579K,2009ApJ...693..112P,2009MNRAS.395..144W,2009MNRAS.393.1302W,2009ApJ...699.1307B,2009arXiv0912.1610C,2010ApJ...709..644I,2009arXiv0909.1968V,2010arXiv1002.3076H,2010arXiv1001.4560L,2010ApJ...708..841W}. However, more definitive clues are required in order to establish a direct cause-and-effect link between major mergers and the appearance of mETGs. 

Studies on the time evolution of the intrinsic properties of mETGs have not provided conclusive results  on their formation and evolution mechanisms yet. While some authors find that the evolution of colors, galaxy populations, number densities, and structural properties of mETGs has been negligible since $z\sim 2$ \citep[directly questioning the hierarchical picture, see][]{2003A&A...402..837P,2006AJ....131.1288B,2007MNRAS.375.1025C,2008ApJ...680...41D,2009MNRAS.396.1573F}, other studies report a noticeable evolution in the physical properties of these galaxies, quite coherent with an assembly mostly driven by major mergers \citep{2004ApJ...600L..11B,2004ApJ...608..752B,2006ApJ...639..644E,2006A&A...453..809I,2007MNRAS.382..109T,2007ApJ...654..858B,2007MNRAS.374..614L,2007ApJS..172..406S,2008ApJ...682..919C,2008ApJ...680...41D,2008ApJ...687...50P,2008ApJ...687L..61B,2008ApJ...688..770F,2009ApJ...696.1554C,2009AJ....138..579K}. 

Another alternative to study the contribution of major mergers to mass assembly is the determination of the evolution of major merger fractions with redshift. Although observational errors make different estimates to disagree by more than a factor of $\sim 2$ \citep[][LSJ09c hereafter]{2006ApJ...652..270B,2005AJ....130.1516D,2009MNRAS.394.1956C,2009A&A...498..379D,2009A&A...497..743H,2009ApJ...697.1971J,2009ApJ...694..643L}, these studies have posed two main results: 1) that the number of major mergers has decreased strongly since $z\sim 1$, and 2) that wet mergers (\ie, involving two gas-rich disks) were more frequent than dry (ETG-ETG mergers) and mixed events (ETG-disk mergers) at $z>0.6$, becoming dry mergers dominant at $z<0.3$ \citep[see][LSJ09a hereafter]{2003ApJ...597L.117K,2008ApJ...679..260M,2008ApJ...683L..17T,2009MNRAS.394.1956C,2009A&A...498..379D,2009MNRAS.396.2003L,2009A&A...501..505L}. 

As the observational properties of the most massive ETGs (mainly, ellipticals) are better reproduced by simulations of dry major mergers than of wet events \citep{2003ApJ...597..893N,2006ApJ...636L..81N,2006ApJ...641...90R,2007MNRAS.379..401E}, some authors conclude that dry mergers at $z<0.3$ must have driven the final assembly of elliptical galaxies  \citep{2005AJ....130.2647V,2009arXiv0911.0044R}. This idea is also supported by the low number detected of blue galaxies bright enough to be the gas-rich progenitors of ellipticals \citep[][]{2004ApJ...608..752B,2007ApJ...665..265F}. However, the significant drop registered in the number density of star-forming galaxies since $z\sim 1$ and the characteristics of post-starburst galaxies entering in the red sequence at $z\sim 0.7$ point more directly to wet than to dry events for explaining the stellar mass migration observed from the blue galaxy cloud to the red sequence in the last $\sim 8$\,Gyr \citep{2006A&A...455..879Z,2007ApJS..172..406S,2010ApJ...709..644I,2009MNRAS.395..144W}. 

Summarizing, while some authors question not only the relevance of dry mergers in the final buildup of the mETGs \citep{2006ApJ...644...54M,2007ApJS..172..494S}, but of all major mergers \citep{2006ApJ...648..268B}, others claim that major merging must have been essential in it \citep{2005ApJ...625..621B,2009ApJ...697.1369B,2006ApJ...640..241B,2007A&A...476..137A,2009MNRAS.395..144W,2010arXiv1001.4560L}. Definitely, the relative role of wet vs.\,dry vs.\,mixed mergers in the recent assembly of mETGs is still unsettled.

In order to reconcile all these observational facts, mixed scenarios for the formation of mETGs have been proposed, in which blue galaxies have their SF quenched in gas-rich mergers, migrate to the red sequence, and merge further through mixed and dry mergers \citep[][]{2007ApJ...665..265F,2008ApJS..175..390H,2008ApJS..175..356H}. A direct verification of the feasibility of this scenario accounting \emph{strictly} for the major mergers reported by observations has not been carried out yet. So, we have approached this question directly through semi-analytical modelling,  studying how the present-day mETGs would have evolved backwards-in-time under the hypothesis that each observed major merger gives place to an ETG. Results are being published in a series of papers. In the first paper of this series (\citealt{2010arXiv1002.3537E}, \citetalias{2010arXiv1002.3537E} hereafter), we showed that it is completely feasible to reproduce the observational buildup of $\sim 50$-60\% of  present-day mETGs just accounting for the effects of the major mergers strictly reported by current observations since $z\sim 1$. In the present-paper (Paper II), we analyse in detail how the coordinated action of wet, mixed, and dry mergers since $z\sim 1$ explains this buildup, showing that many observational results that are apparently against the hierarchical scenario can be reconciled with it. 

The present paper is organized as follows. In \S\ref{sec:model}, we give a brief outline of the model. Section \S\ref{sec:results} is devoted to the presentation of results. In \S\ref{sec:numberevolution}, we analyse the model predictions on the number evolution experienced by mETGs at $z\lesssim 1$ through major mergers. The relative role of each merger type in the recent buildup of mETGs is analysed in \S\ref{sec:roleofmergers}. Section \S\ref{sec:comparison} compares the model predictions with different observational and theoretical estimates at different redshifts. In \S\ref{sec:gaspoormergers}, we quantify the contribution of gas-poor mergers (dry and mixed) to the recent mETGs buildup. Section \S\ref{sec:mETGinplace} discusses the model predictions on the fraction of present-day mETGs that are really in place since $z\sim 1$. The discussion of results and a brief summary of them can be found in \S\S\ref{sec:discussion} and \ref{sec:conclusions}, respectively. We will use a $\Omega_M = 0.3$, $\Omega_\Lambda = 0.7$, $H_0 = 70$ km s$^{-1}$ Mpc$^{-1}$ concordant cosmology throughout the paper. All magnitudes are given in the Vega system.

\section{The model}
\label{sec:model}

A brief outline of the model is given in this section. For a more detailed model description, the reader is referred to \citetalias{2010arXiv1002.3537E}.

\subsection{Basic outline}
\label{sec:basic}

The model adopts the backwards-in-time technique first introduced by \citet{1980ApJ...241...41T}. It traces back-in-time the evolution of the local galaxy populations considering two different sources of evolution: the number evolution that observational merger fractions imply at each redshift, and the typical luminosity evolution of each galaxy type due to its star formation history (SFH). The evolution of the volume element for the considered cosmology is also taken into account. The novelty of the model is the realistic treatment of the effects of major mergers on the LFs of galaxy populations (see \S\ref{sec:merging}). 

The model is based on the \ncmod\ code, created by \citet[][G98 hereandafter]{1998PASP..110..291G} for generating predictions on galaxy number counts. The code computes the L-evolution experienced by each galaxy type, according to its SFH. The following local morphological types have been considered: E-S0a, Sa-Sab, Sb-Sbc, Sc-Scd, and Sd-Irr \citepalias[see][for more information]{2010arXiv1002.3537E}. We have used standard parametrizations for the SFH of each galaxy type, and metallicities and characteristic optical depths for each morphological type according to observations. For all galaxy types, the SF starts at $z_\mathrm{f,*}= 3$, where the SFH of the Universe exhibits its peak \citep{2006ApJ...651..142H}. Notice that this redshift represents the epoch at which the bulk of SF starts in the Universe, that can be different to the redshift at which a galaxy is completely assembled, because of mergers \citep{2006MNRAS.366..499D}. 

We have implemented several improvements to the original code. In particular, the traditional counting method and the procedures dealing with the L-evolution and number-evolution have been modified. The original spectral evolutionary library of the code has also been updated to {\tt GALAXEV\/}, the isochrone synthesis code by \citet[][]{2003MNRAS.344.1000B}. 

The original merging procedure implemented by G98 simply removes a given fraction of galaxies at each redshift from those galaxy types that the user considers to be affected by mergers. In this regard, the procedure is undoing the remnant galaxies into their original progenitors (we will refer to this as \emph{reversed merging}). Nevertheless, the original code implementation provides a too simplistic view of their role in galaxy evolution: it does not consider the dramatic morphological transformations that observational major mergers usually drive, neither the typical duration of a major merger (real major mergers are not instantaneous), nor the characteristics of the transitory, intermediate phases undergone by the progenitor galaxies during a major merger. For these reasons, a more realistic merging procedure, strictly based on robust observational and computational results, has been implemented in this study (see \S\ref{sec:merging}). 

\subsection{Major mergers procedure}
\label{sec:merging}

The main hypothesis of the model is to consider that the number density of ETGs that are assembled through major mergers at each redshift $z$ is equal to the number density of major mergers that are observed at that redshift; \ie, we are implicitly assuming that the final remnant of each major merger is an ETG, as posed by N-body simulations (see references in \S\ref{sec:introduction}).

This assumption does not contradict the disk rebuilding scenario proposed by \citet{2005A&A...430..115H}. According to it, a large disk could be rebuilt after a major merger in $\sim 2$-3\,Gyr, in the case that the progenitor galaxies contain a large gas reservoir  \citep[amounting to $\sim 50$\% of their masses, see][]{2009arXiv0903.3962H,2009arXiv0903.3961P}. This scenario is supported by observational examples at $0.6<z<1$  \citep{2005A&A...430..115H,2009MNRAS.398..312G,2009A&A...496..381H,2009A&A...493..899P,2009A&A...501..437Y} and by numerical and cosmological simulations \citep[see numerous references in the introduction by][]{2009ApJ...702.1005S}. However, a close inspection to the sufficiently-relaxed observational examples shows that those cases related with major mergers are basically S0-S0a galaxies, whereas those exhibiting lower bulge-to-disk ratios (and thus, not being ETGs according to our definition) are associated to minor mergers instead \citep[][]{2006A&A...457...91E,2008A&A...483L..39C,2009ApJ...693..112P}. Therefore, the disk rebuilding does not conflict with our main hypothesis at all \citep[see an interesting discussion on the topic in][]{2009arXiv0911.1126O}. 

The model does not consider the effects of minor mergers. Major mergers drive much more dramatic structural changes, violent starbursts, and higher mass increments in a galaxy than minor mergers \citep{1991ApJ...370L..65B,2000MNRAS.316..315B,2003ApJ...597..893N,2005MNRAS.357..753G,2006A&A...457...91E,2006MNRAS.372L..78G,2007A&A...476.1179B}. Even considering that minor mergers have probably been more numerous than major mergers \citep[at least, by a factor of $\sim 2$, see \eg,][]{2006MNRAS.370.1905H,2008IAUS..245...63C,2008ASPC..396..243K,2008ApJ...683..597S,2009ApJ...702.1005S,2009ApJ...702..307S,2009ApJ...697.1971J,2009MNRAS.394.1713K,2009ApJ...699L.178N}, observations indicate that their contribution has been significant only in low-mass systems \citep[with $\mathcal{M}_*/\Msun <10^{10}$, see][]{2009ApJ...697.1369B,2009arXiv0911.1126O,2010ApJ...710.1170L}. 

In the model, we have considered the number of major merger events that are derived at each redshift from the observational merger fractions obtained by \citetalias{2009A&A...501..505L}, computed at $z\lesssim 1$ for galaxies with $M(B) < -20$\,mag through asymmetry methods. \citetalias{2009A&A...501..505L} estimates have the advantage of being more robust than others, because they account for the uncertainties due to errors in redshift determinations and to the inherent loss of morphological information at higher redshifts \citep[][]{2008PASP..120..571L}. The model predictions are limited by the assumed merger fractions in two aspects: 1) results can only be derived for galaxies brighter than $M(B) = -20$\,mag at all redshifts, and 2) results are only reliable where these merger fractions have been computed, \ie, at $z \lesssim 1$. Results at $z>1$ are derived using an extrapolation of \citetalias{2009A&A...501..505L} major merger fractions at $z<1$, so they must be considered cautiously \citepalias[see the comments on the model limitations at $z>1$ in][]{2010arXiv1002.3537E}. 

As the model evolves the local galaxy populations backwards-in-time, major mergers occurring at each redshift are reversed, in the sense that each ETG coming from a major merger at a certain redshift is decomposed into its two progenitor galaxies at earlier epochs. As major mergers tend to occur only between very-late or very-early types \citep{2005AJ....130.1516D,2009A&A...503..379T,2009A&A...508.1217Z}, the progenitor galaxies of a major merger are assumed to be of Sd-Irr type if it is a gas-rich progenitor, or an ETG if it is a gas-poor one. 

Observationally, the conversion of two merging galaxies into a remnant ETG is not instantaneous. So, we have considered the different phases, timescales, and colors of progenitor galaxies during a major merger, as stated by observations and N-body simulations for wet, mixed, and dry mergers \citep{2008MNRAS.384..386C,2008MNRAS.391.1137L,2010MNRAS.401.1043D}. The number density of each merger type have been computed at each redshift using the relative fractions of wet, mixed, and dry mergers derived observationally by \citet{2008ApJ...681..232L} at $z\lesssim 1.2$. For more details, consult \citetalias{2010arXiv1002.3537E}.

\section{Results}
\label{sec:results}

In \citetalias{2010arXiv1002.3537E}, we showed that the model reproduces the evolution of the galaxy LFs up to $z\sim 1$, simultaneously for different bands ($B$, $I$, $K$) and selection criteria (on color, on morphology). The model also proves the feasibility of building up $\sim 50$-60\% of the present-day mETGs at $z\lesssim 1$ through the major mergers strictly reported by observations, in general agreement with mass-downsizing trends. In the present paper, we analyse in detail how the model explains the recent buildup of mETGs through the coordinated action of wet, mixed, and dry mergers since $z\sim 1$.

As indicated above, the model results are valid for galaxies with $M(B)\leq -20$\,mag at $z\lesssim 1$, the limiting magnitude imposed by \citetalias{2009A&A...501..505L} merger fractions. This magnitude corresponds roughly to $M\lesssim M^{\ast}$ galaxies at $z\lesssim 1$, or to ETGs at $z\lesssim 1$ with masses $\mathcal{M}_{*}>10^{11}\Msun$ at $z=0$ \citep{2006A&A...453L..29C}. In order to derive the results of the present paper, we have considered strictly the mETGs at each redshift that will have $\mathcal{M}_{*}>10^{11}\Msun$ at $z=0$ from the k-corrected and e-corrected $K$-band LFs predicted by the model. The effects of the dust-extinction and the L-evolution have also been taken into account. Model uncertainties due to the observational errors of \citetalias{2009A&A...501..505L} merger fractions and of the relative fractions of wet, mixed, and dry major mergers derived by \citet{2008ApJ...681..232L} at each redshift have been considered for determining the error bars and uncertainty regions plotted in all the figures \citepalias[see][for more information on model uncertainties]{2010arXiv1002.3537E}.

\begin{figure}[t]
\begin{center}
\includegraphics*[width=0.5\textwidth,angle=0]{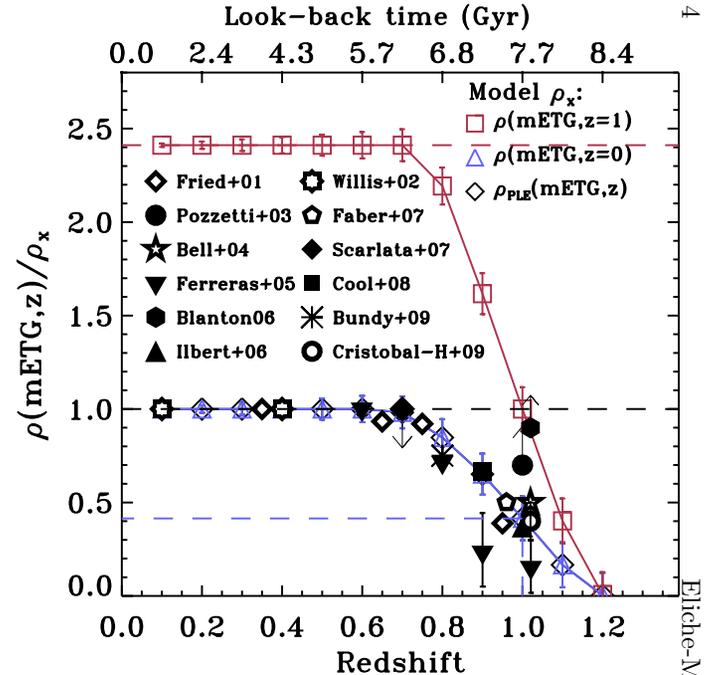}
\caption{Evolution of the number density of mETGs with each redshift, with respect to the number density of mETGs at the same redshift in PLE (\emph{open diamonds}), to the number of mETGs at $z=1$ derived in the model (\emph{open squares}), and to the present-day number of mETGs (\emph{open triangles}). Error bars account for the observational errors in the used merger fractions and in the relative percentages of each merger type with redshift. \emph{Rest of symbols}: Observational estimates (consult the legend in the figure). Error bars on top of observational estimates indicate the range reported by a certain author. \emph{[A color version of this plot is available at the electronic edition.]}
} \label{fig:etgnumbergrowth}
\end{center}
\end{figure}

\subsection{Number evolution of mETGs since $z\sim 1$}
\label{sec:numberevolution}

Figure\,\ref{fig:etgnumbergrowth} represents the predicted evolution of the number density of mETGs with redshift up to $z\sim 1.2$. The figure shows the fraction of mETGs existing at each redshift, with respect to three different quantities: 1) to the number density of mETGs at the same redshift in a pure luminosity evolution (PLE) model (\ie, in a model identical to ours, but without major mergers), 2) to the number density of mETGs at $z=1$, as derived by the model, and 3) to the present-day number density of mETGs (open diamonds, squares, and triangles, respectively). Obviously, cases 1 and 3 are equivalent, as the number density of mETGs in a PLE model does not change with redshift, being constantly equal to the number density at $z=0$. 

The number density of mETGs predicted by the model at $z\sim 1$ is $\sim 40(^{50}_{30})$\% of its present-day value\footnote{Hereafter, the values in superscript and subscript format refer to the upper and lower limits of the uncertainty region around the nominal model prediction, due to the errors in the observational merger fractions by \citetalias{2009A&A...501..505L} and in the relative percentages of each merger type by \citet{2008ApJ...681..232L}.} (diamonds in the figure). Therefore, $\sim 60(^{70}_{50})$\% of the present-day number density of mETGs has appeared into the cosmic scenario  since $z\sim 1$. 

We have overplotted the estimates derived at different redshifts by several observational studies in the figure \citep[][]{2001A&A...367..788F,2002MNRAS.337..953W,2003A&A...402..837P,2004ApJ...608..752B,2005ApJ...635..243F,2006ApJ...648..268B,2006A&A...453..809I,2007ApJ...665..265F,2007ApJS..172..406S,2008ApJ...682..919C,2009ApJ...696.1554C}. The model reproduces the observed evolution of the number density of mETGs with redshift pretty well up to $z\sim 1$, demonstrating that the numerical appearance of $\sim 60$\% of the present-day mETGs can be explained just considering the major mergers reported by current observations up to $z\sim 1$ \citepalias[see][]{2010arXiv1002.3537E}. Moreover, the model predicts naturally that most of this numerical evolution takes place during the interval of $\sim 1.4$\,Gyr elapsed at $0.7<z<1$, in agreement with most observations. 

This appearance of $\sim 60$\% of the present number density of mETGs since $z\sim 1$ means that the number density of mETGs at $z=1$ has increased by a factor of $\sim 2.5$ in the last $\sim 8$\,Gyr, basically between redshifts $z=1$ and $z\sim 0.7$ (empty squares in the figure). However, this numerical growth of mETGs is frozen at lower redshifts: since $z\sim 0.7$, the number density of mETGs remains nearly constant, accordingly with observations. We will show that the model is capable of reproducing this frostbite just accounting for the decrease of the major merger fractions with redshift that observations report (see \S\ref{sec:roleofmergers}). So, the model suggests that major merging basically drives the evolution of the stellar mass density since $z\sim 1$ in high-mass systems, in agreement with several authors  \citep{2005ApJ...625..621B,2007A&A...476..137A,2009A&A...498..379D}, although it shows that their effects have been noticeable at epochs earlier than $z\sim 0.7$.

\begin{figure}[!]
\begin{center}
\includegraphics*[width=0.5\textwidth,angle=0]{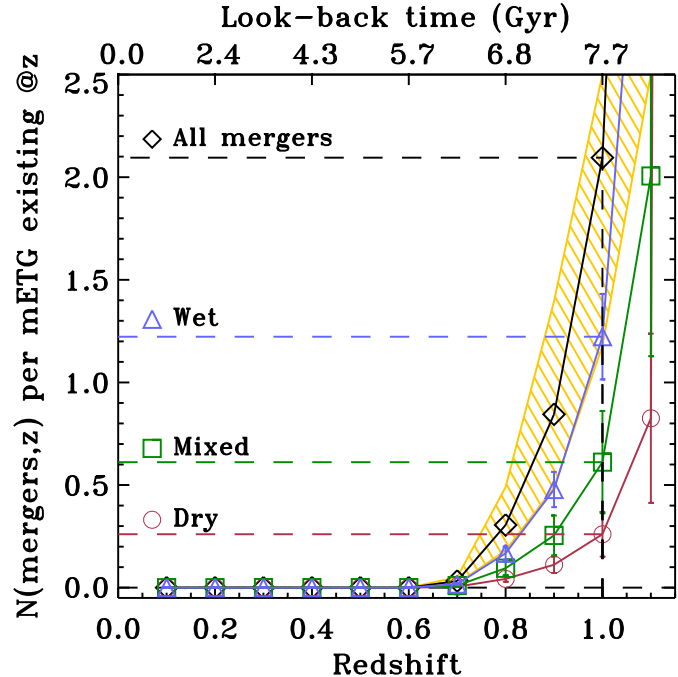}
\caption{Number of wet, mixed, dry, and all major mergers occurring at each redshift, per mETG existing at that redshift (\emph{triangles}, \emph{squares}, \emph{circles}, and \emph{diamonds}, respectively). The uncertainty regions (dashed area) and the error bars account for the observational errors of the used merger fractions and of the assumed relative percentages of each merger type with redshift. \emph{[A color version of this plot is available at the electronic edition.]}
}\label{fig:nmergerperetg}
\end{center}
\end{figure}

\subsection{Relative role of dry, mixed, and wet major mergers in the buildup of mETGs at $z\lesssim 1$}
\label{sec:roleofmergers}

In Fig.\,\ref{fig:nmergerperetg}, we show the predicted number of wet, mixed, and dry major mergers occurring at each redshift, per mETG existing at that redshift (triangles, squares, and circles, respectively). The predicted number of major mergers at $z\sim 1$ per mETG at this redshift is $\sim 2.1(^{4.5}_{1.2})$ (diamonds). Nearly half of these major mergers were wet, \ie, $\sim 1.2$ mETG has appeared through a wet merger into the cosmic scenario at $z\sim 1$ per mETG existing at that redshift. So, the population of mETGs at $z\sim 1$ was being doubled at that redshift just accounting for wet major mergers. The contribution of mixed and dry mergers at $z\sim 1$ is also significant: the number of mixed major mergers per mETG at $z\sim 1$ is nearly $\sim 0.60(^{0.90}_{0.30})$, and $\sim 0.25(^{0.35}_{0.15})$ for dry events. This means that nearly all mETGs at $z\sim 1$ ($\sim 85$\%) have taken part as gas-poor progenitors in a major merger at $0.9<z<1$.

The number of major mergers per mETG at each redshift decreases quickly to lower redshifts, being the effects of all merger types completely negligible at $z\lesssim 0.7$, compared to the population of mETGs at those redshifts. Then, although the model poses that major mergers must have driven the appearance of $\sim 60$\% of the present-day number density of mETGs into the cosmic scenario at $0.7<z<1$, it also indicates that their contribution to it has been insignificant during the last $\sim 6.3$\,Gyr in general (this is why the number density of mETGs has remained nearly constant since $z\sim 0.7$, see \S\ref{sec:numberevolution}).

\begin{figure}[!]
\begin{center}
\includegraphics*[width=0.5\textwidth,angle=0]{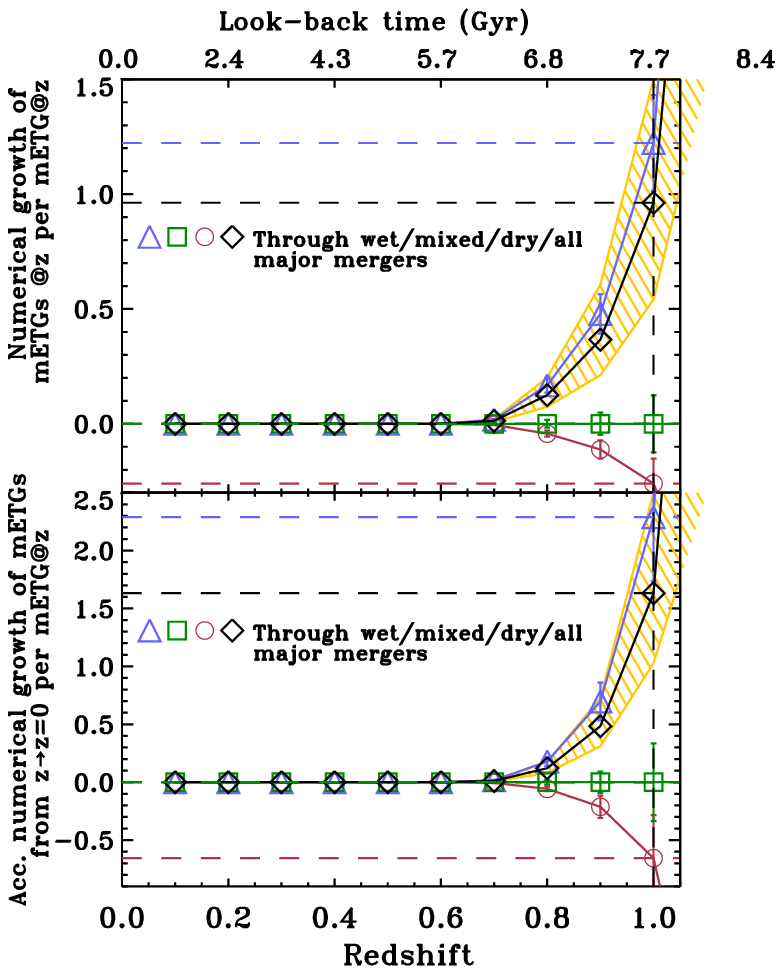}
\caption{Predicted numerical growth of mETGs driven by each type of major merger at each redshift (\emph{upper panel}), and accumulated since that redshift down to $z=0$ (\emph{lower panel}), per mETG existing at that redshift (\emph{triangles}, \emph{squares}, and \emph{circles} for wet, mixed, and dry major mergers, respectively). \emph{Diamonds}: Net numerical growth of mETGs driven by the coordinated action of all major mergers. \emph{[A color version of this plot is available at the electronic edition.]}
}\label{fig:etggrowthratez}
\end{center}
\end{figure}

In the upper panel of Fig.\,\ref{fig:etggrowthratez}, we have plotted the predicted net numerical growth of mETGs driven by each merger type (wet, mixed, and dry) at each redshift, per mETG existing at that redshift. This net numerical growth indicates the net number of mETGs that are appearing/disappearing into the cosmic scenario through each merger type. Notice that, besides the galaxy transfers across mass bins due to the SF or to the mass increase due to mergers, the effect of a wet merger on the number density of mETGs is the appearance of a "brand-new" mETG in the Universe, whereas dry mergers imply the net decrease of this number in one mETG. Mixed mergers are going to imply a nearly null numerical growth of mETGs in the model, as one ETG merges with a gas-rich galaxy to generate another, more massive ETG. Accounting for this, we have also overplotted the net numerical growth of mETGs driven by the combined effects of all major mergers at each redshift (diamonds in the figure). 

We corroborate again that the role of major mergers has been relevant at $0.7<z<1$: only at $z\sim 1$, major mergers are responsible of creating one new mETG per mETG already existing at that redshift. Most of this growth is driven by wet mergers, but the effects of dry mergers are extremely relevant, as the net numerical growth of mETGs at $z\sim 1$ would be higher by more than $\sim 25$\% of its actual value if dry mergers did not exist.

In the lower panel of Fig.\,\ref{fig:etggrowthratez}, we show the same as in the upper panel, but accumulated since a given redshift down to $z=0$. The model predicts that major mergers have built up $\sim 1.6(^{2.3}_{1.2})$ new ETGs per mETG existing at $z\sim 1$ in the last $\sim 8$\,Gyr (diamonds), \ie, the number density of mETGs at $z\sim 1$ has increased by a factor of $\sim 2.5$ since then through major mergers (in agreement with the result stated at \S\ref{sec:numberevolution}). The major driver of this growth are wet major mergers, which could have created $\sim 2.3(^{2.9}_{1.7})$ mETGs per mETG already existing at $z\sim 1$ since that epoch (triangles). However, dry mergers have been responsible of removing $\sim 0.6(^{0.9}_{0.3})$ mETGs per mETG already existing at $z\sim 1$ at the same time (circles in the figure). 

As indicated above, the bulk of mETG assembly through major mergers is occurring at $z>0.7$. At lower redshifts, the decrease of the major merger fractions that observations report is enough to explain the observed "frostbite" in the generation of mETGs at $z<0.7$ (in terms of net numerical growth). Note that the effect of a higher relative relevance of dry and mixed mergers than of wet mergers at $z<0.8$ \citep{2008ApJ...681..232L} is secondary and negligible in driving this frostbite. In fact, even if all the major mergers at $z<0.7$ were wet, they would still imprint a negligible change to the existing number density of mETGs (see Fig.\,\ref{fig:etggrowthratez}). 

Moreover, the model predicts specifically that dry major mergers are irrelevant as compared to the population of mETGs existing at $z<0.7$ (circles in the lower panel of Fig.\,\ref{fig:etggrowthratez}). This prediction agrees with studies supporting evidence against a large contribution of dry mergers to the formation of massive spheroidal galaxies at $z < 0.4$ \citep{2006ApJ...644...54M,2007ApJS..172..494S}. In fact, if dry major mergers were determinant at $z<0.7$ for the buildup of ETGs with $\mathcal{M}_*/\Msun>10^{11}$, observations should detect a significant decrease in their number density at $z<0.6$ that is not observed (see Fig.\,\ref{fig:etgnumbergrowth}). 

\begin{figure}[t]
\begin{center}
\includegraphics*[width=0.5\textwidth,angle=0]{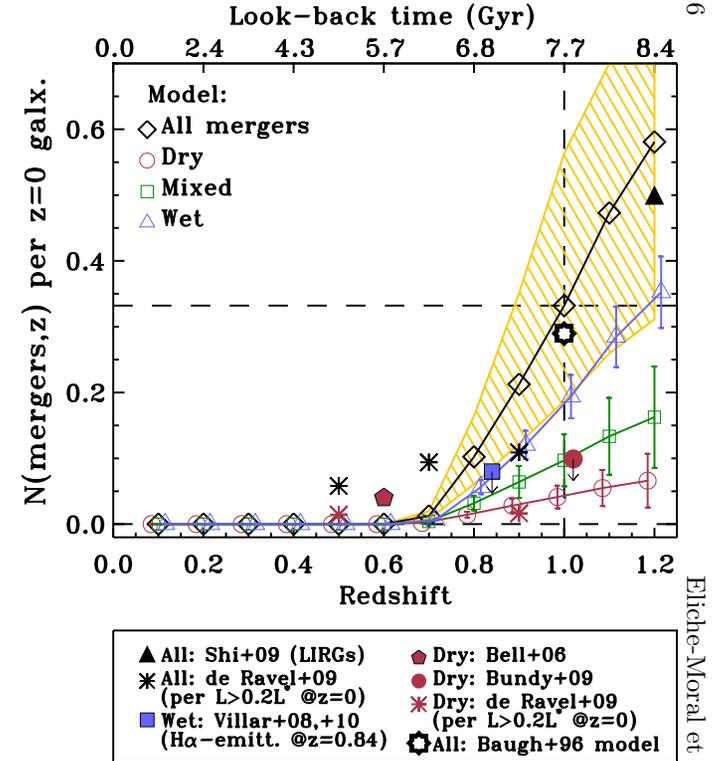}
\caption{Predicted average number of wet, mixed, dry, and all major merger events per present-day $L>L^*$ galaxy occurred at each redshift bin $[z-0.1,z]$, for $z\lesssim 1.2$. \emph{Open diamonds}: All major mergers. \emph{Open triangles}: Wet mergers. \emph{Open squares}: Mixed mergers. \emph{Open circles}: Dry mergers. \emph{Remaining symbols}: Observational estimates derived by different studies (consult the legend in the figure). \emph{[A color version of this plot is available at the electronic edition.]}
}\label{fig:nmergerstot1}
\end{center}
\end{figure}

\begin{figure}[t]
\begin{center}
\includegraphics*[width=0.5\textwidth,angle=0]{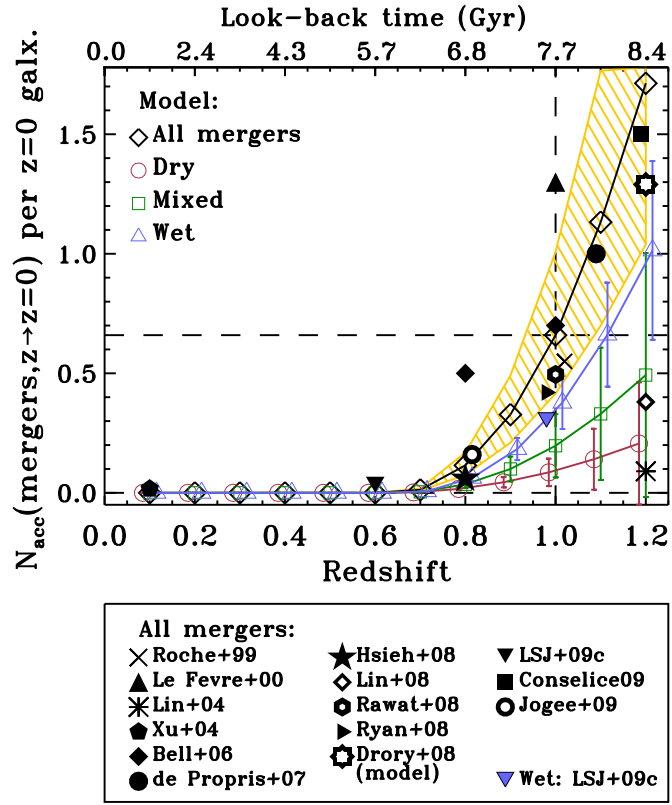}
\caption{Predicted average number of wet, mixed, dry, and all major merger events per present-day $L>L^*$ galaxy occurred since each redshift $z$ down to the present, for $z\lesssim 1.2$. \emph{Open diamonds}: All major mergers. \emph{Open triangles}: Wet mergers. \emph{Open squares}: Mixed mergers. \emph{Open circles}: Dry mergers. \emph{Remaining symbols}: Observational estimates derived by different studies (consult the legend in the figure). \emph{[A color version of this plot is available at the electronic edition.]}
}\label{fig:nmergerstot2}
\end{center}
\end{figure}

\subsection{Comparison with observations}
\label{sec:comparison}

\subsubsection{Number of major mergers per local $L>L^{*}$ galaxy}
\label{sec:comparison1}

In this section, we compare the predictions of our model with several estimates derived from observational and theoretical studies. Figure\,\ref{fig:nmergerstot1} shows the predicted average number of major merger events occurred per local $L\gtrsim L^*$ galaxy at each redshift, for wet, mixed, dry, and all major mergers, at each redshift bin $[z-0.1,z]$. Several estimates from observations and simulation studies have been overplotted for comparison \citep[][2010 in preparation]{1996MNRAS.283.1361B,2006ApJ...652..270B,2009ApJ...697.1369B,2009A&A...498..379D,2009arXiv0903.3035S,2008ApJ...677..169V}. Note that special care must be taken if this figure is interpreted as the number of major mergers undergone per local galaxy, as not all types of local galaxies are susceptible of deriving from a major merger (see \S\ref{sec:merging}).

From this figure, we can derive that the amount of galaxies involved in a major merger at $z\sim 1$ is equivalent to $\sim 35(^{55}_{20})$\% of the present-day number of $L\gtrsim L^*$ galaxies (empty diamonds). Nearly 5\% correspond to dry mergers (empty circles), $\sim 10$\% to mixed mergers (empty squares), and $\sim 20$\% to wet mergers (empty triangles). These percentages can vary by less than $\sim 15$\% of their values due to observational errors. The number of mergers undergone at each redshift is predicted to decrease quickly for all merger types between $z\sim 1.0$ and $z\sim 0.7$, becoming nearly negligible at lower redshifts as compared to the local population of $L>L^*$ galaxies. 

The model provides a good match to the number of mergers per present-day $L>L^{*}$ galaxy derived at $z\sim 1.2$ from luminous infrared galaxies by \citet{2009arXiv0903.3035S}. Moreover, our prediction of the number of wet mergers at $z\sim 0.8$ per local $L>L^{*}$ galaxy coincides pretty well with the number of H$\alpha$-emitting galaxies undergoing a merger at $z\sim 0.84$, most of which are likely undergoing a gas-rich merger \citep[see][2010 in preparation]{2008ApJ...677..169V}. However, the model predicts a lower number of dry mergers than the studies by \citet{2006ApJ...652..270B} at $z\sim 0.6$ and \citet{2009ApJ...697.1369B} at $z\sim 1$. Although not directly comparable with our predictions, the observational estimates for $L>0.2L^*$ galaxies obtained by \citet{2009A&A...498..379D} have also been overplotted, in order to remark the strong dependence that merger fractions exhibit on the galaxy mass \citep[for more information, see][]{2009ApJ...697.1369B,2009A&A...501..505L}. 

In Fig.\,\ref{fig:nmergerstot2}, we compare the predicted number of major mergers of each type occurred since each redshift down to the present per local $L>L^*$ galaxy with several observational and theoretical estimates \citep[][LSJ09c]{1999MNRAS.306..538R,2000MNRAS.311..565L,2004ApJ...617L...9L,2008ApJ...681..232L,2004ApJ...603L..73X,2006ApJ...652..270B,2007ApJ...666..212D,2008ApJ...680...41D,2008ApJ...683...33H,2008ApJ...681.1089R,2008ApJ...678..751R,2009MNRAS.394.1956C,2009ApJ...697.1971J}. The number of major mergers occurred since $z\sim 1$ amounts to $\sim 65(^{110}_{40})$\% of the present number of $L>L^*$ galaxies. According to the model, there have been $\sim 0.35$ wet major mergers, $\sim 0.2$ mixed mergers, and only $\sim 0.1$ dry mergers per local $L>L^*$ galaxy. The errors in the observational relative percentages of each merger type affects negligibly to the previous estimates (see the error bars in the figure). 

The model predictions are in excellent agreement with observational estimates at different redshifts up to $z\sim 1.2$ \citep[][LSJ09c]{1999MNRAS.306..538R,2007ApJ...666..212D,2008ApJ...680...41D,2008ApJ...683...33H,2008ApJ...681.1089R,2008ApJ...678..751R,2009MNRAS.394.1956C,2009ApJ...697.1971J}. Moreover, the prediction for specifically wet mergers at $z\sim 1$ coincides pretty well with the estimate by \citetalias{2009ApJ...694..643L} from the GOYA Survey \citep{2007RMxAC..29..165A}. Only observational estimates derived from close pair companions or two-point correlation functions differ noticeably from the model predictions \citep[see data points by][in the figure]{2006ApJ...652..270B,2004ApJ...617L...9L,2008ApJ...681..232L}. It should be noticed also that the observational estimates by themselves are posing a negligible contribution of major mergers at $z<0.7$ (see data points in the figure). So, our model demonstrates that the frostbite of the generation of new mETGs since $z\sim 0.7$ can be explained just accounting for the observational decrease of the major merger fraction registered since $z\sim 1$ to the present.

\begin{figure}[t]
\begin{center}
\includegraphics*[width=0.5\textwidth,angle=0]{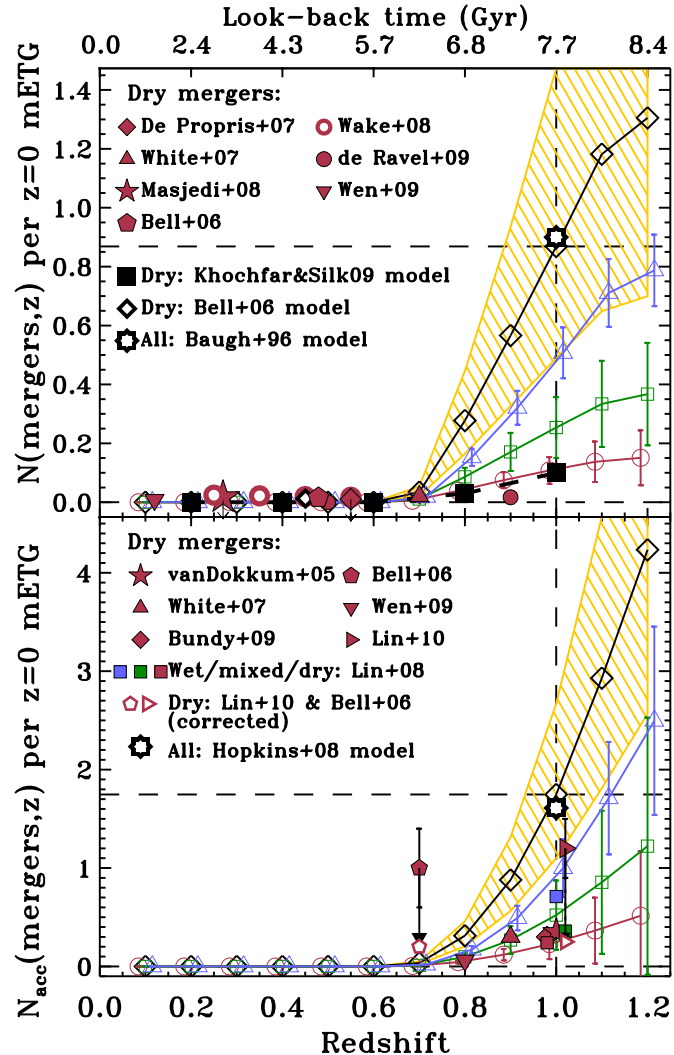}
\caption{Predicted average number of wet, mixed, dry, and all major merger events per present-day mETG occurred at each redshift bin $[z-0.1,z]$ (\emph{top panel}), and accumulated between redshift $z$ and $z=0$ (\emph{bottom panel}), up to $z\sim 1.2$. \emph{Open diamonds}: All major mergers. \emph{Open triangles}: Wet mergers. \emph{Open squares}: Mixed mergers. \emph{Open circles}: Dry mergers. \emph{Remaining symbols}: Observational estimates derived by different studies (consult the legend in the figure). The original observational estimates by \citet{2010arXiv1001.4560L} and \citet{2006ApJ...640..241B} in the lower panel assume that the merger rate is constant with redshift. Accounting for the real redshift-dependence of merger fractions, both estimates agree pretty well with our model predictions and with the remaining observational values (consult the legend). The plotted estimate from the \citet{2008ApJS..175..390H} model corresponds to ETGs with local masses $\log(\mathcal{M}_*/\Msun)\sim 11$. \emph{[A color version of this plot is available at the electronic edition.]}}\label{fig:nmergers}
\end{center}
\end{figure}

\subsubsection{Number of major mergers per local mETG}
\label{sec:comparison2}

Figure\,\ref{fig:nmergers} is similar to Figs.\,\ref{fig:nmergerstot1}-\ref{fig:nmergerstot2}, but refers the number of major merger events (wet, mixed, dry, and total) now to the present-day number of mETGs. The upper panel of the figure shows the predicted number of major mergers occurred at each redshift bin $[z-0.1,z]$, per local mETG. Observational and theoretical estimates of the number of dry major mergers undergone per present-day mETG at different redshifts have been overplotted for comparison in this panel \citep[][]{1996MNRAS.283.1361B,2006ApJ...640..241B,2007ApJ...666..212D,2007ApJ...655L..69W,2008ApJ...679..260M,2008MNRAS.387.1045W,2009A&A...498..379D,2009MNRAS.397..506K,2009arXiv0909.3365W}. The agreement between the model predictions and observations and estimates of hierarchical models is quite good at $z\lesssim 1$. 

According to our model, the number of major mergers per present-day mETG at $z=1$ was $\sim 0.85(^{1.50}_{0.50})$ (diamonds in the upper panel of the figure). This fraction decreases quickly, being $\lesssim 10$\% at $z<0.7$. At $z\sim 1$, nearly 0.55 major mergers per local mETG corresponds to wet events (empty triangles), while the rest of mergers involve a gas-poor galaxy ($\sim 0.23$ are mixed --empty squares--, and $\sim 0.07$ are dry --empty circles). These fractions decay strongly at lower redshifts. 

The bottom panel of Figure\,\ref{fig:nmergers} represents the accumulated number of major mergers of each type occurred since each redshift $z$ to the present, per present-day mETG. Different observational estimates have been overplotted for comparison again \citep[][]{2005AJ....130.2647V,2006ApJ...640..241B,2007ApJ...655L..69W,2008ApJS..175..390H,2008ApJ...681..232L,2010arXiv1001.4560L,2009ApJ...697.1369B,2009arXiv0909.3365W}. The predictions are in good general agreement with most of the consulted observational estimates, particularly with respect to dry mergers (see the figure). So, we can conclude that the model predicts a realistic number of wet, mixed, and dry major mergers per local $L\gtrsim L^*$ galaxy and per local mETG at different redshifts at $z\lesssim 1.2$.

According to the model, there have been $\sim 1.8(^{2.6}_{1.2})$ major mergers since $z\sim 1$ per local mETG (diamonds in the lower panel of the figure). Approximately 1 of these $\sim 2$ mergers takes place during a period of $\sim 0.5$\,Gyr elapsed at $0.9<z<1$, while the other one basically occurs during the period of $\sim 1$\,Gyr elapsed at $0.7<z<0.9$. Of these $\sim 2$ mergers, $\sim 0.35$ correspond to dry mergers (empty circles) and $\sim 0.6$ to mixed mergers (empty squares). This means that $\sim 0.95$ of the $\sim 2$ major mergers occurred in the last $\sim 8$\,Gyr per local mETG was a gas-poor event, being the another one a wet merger (empty triangles). This implies that $\sim 95$\% of local mETGs have experienced a gas-poor event since $z\sim 1$, meaning that most of local mETGs derive from at least another ETG which has undergone a major merger in the last $\sim 8$\,Gyr. The remaining local mETGs not involved in any gas-poor merger since $z\sim 1$ ($\sim 5$\%) could have been in place since $z\sim 1$ or not, in the case that they have been assembled through a wet merger. Therefore, the model indicates that the fraction of local mETGs that have not experienced any major merger since $z\sim 1$ is negligible ($\lesssim 5$\%, see  \S\S\ref{sec:gaspoormergers}-\ref{sec:mETGinplace} for more information).

At the most, $\sim 35$\% of local mETGs have been involved in a dry major merger during the last $\sim 8$\,Gyr. This prediction is an upper limit because ETGs involved in gas-poor mergers could have undergone more than one major merger during their lifetimes (see \S\ref{sec:mETGinplace}). This prediction is significantly lower than the one derived by the \citet{2003ApJ...597L.117K} model\footnote{Their definition of "elliptical galaxy" is roughly equivalent to our definition of ETG, although they select $M(B)<-18$ galaxies, instead of $M(B)<-20$ galaxies, as in our case.}, as they report that $\sim 50$\% of local mETGs may have experienced dry mergers in their recent lifetimes. 

Figure\,\ref{fig:nmergers} poses that wet major mergers must have built up $\sim 1$ mETG per local mETG since $z\sim 1$, stressing the relevant role played by gas-rich mergers in the recent building up of mETGs. This implies that, just considering the effects of wet mergers and the model assumption that each major merger gives place to an ETG, no local mETG should be in place at $z\sim 1$ (all of them should be decomposed into their gas-rich progenitors by that epoch). Nevertheless, the model derives that the number density of mETGs at $z\sim 1$ was already $\sim 40$\% of the local one, as it is observed. Then, how does the model derive such a high number of mETGs at $z\sim 1$? The key point to answer this question is to account for the effects of dry and mixed mergers too (see \S\ref{sec:gaspoormergers}). 

\begin{figure}[!]
\begin{center}
\includegraphics*[width=0.5\textwidth,angle=0]{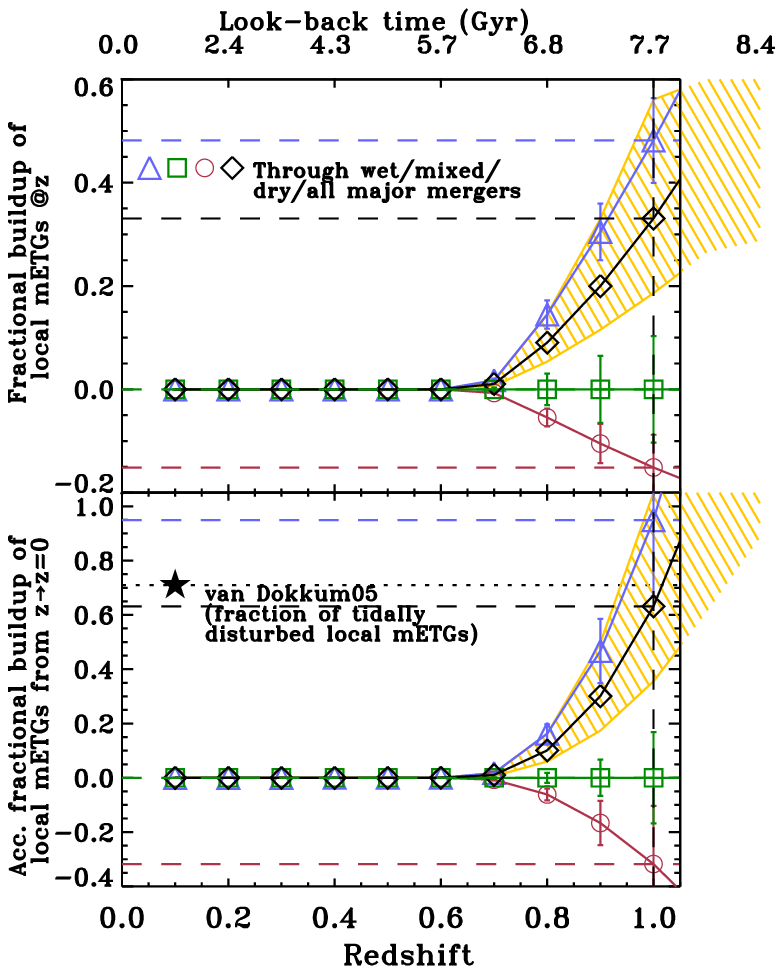}
\caption{
Predicted fractional buildup of the present-day number density of mETGs through major mergers of each type at each redshift (\emph{upper panel}), and accumulated since each redshift down to $z=0$ (\emph{lower panel}). \emph{Symbols}: Fractional increment of the number density of mETGs driven by wet (\emph{triangles}), mixed (\emph{squares}), and dry (\emph{circles}) major mergers. \emph{Empty diamonds}: Net numerical growth of mETGs driven by the coordinated action of all major mergers. \emph{Star}: Fraction of local mETGs exhibiting tidal distortions indicative of a recent major merger, as derived by \citet{2005AJ....130.2647V}. \emph{[A color version of this plot is available at the electronic edition.]}
}\label{fig:etggrowthrate}
\end{center}
\end{figure}

\subsection{The key role of gas-poor major mergers at $0.7<z<1$}
\label{sec:gaspoormergers}

In \S\ref{sec:comparison}, we have shown that nearly $35$\% of present-day mETGs could have experienced a dry major merger since $z\sim 1$ (circles in the bottom panel of Fig.\,\ref{fig:nmergers}), and $\sim 60$\% a mixed merger (see squares in the same panel). Only accounting for the progenitor ETGs of the present-day mETGs involved in the mixed and dry major mergers occurred since that epoch, the mETG population at $z\sim 1$ should be $\sim 130$\% of the present-day mETG population ($\sim 35$\%$\times2+60$\%). Adding the number of mETGs not involved in a gas-poor merger since then ($\sim 5$\% of local mETGs, see \S\ref{sec:comparison}), the mETG population at $z\sim 1$ should be $\sim 135$\% of the present-day one. Additionally, $\sim 95$\% of present-day mETGs have derived from a wet major merger since $z\sim 1$ (triangles in the bottom panel of the figure). This means that the amount of mETGs at $z\sim 1$ must be reduced in $\sim 95$\% of the local number density of mETGs, because the mETGs coming from a wet merger at $z\lesssim 1$ must be decomposed into their two gas-rich progenitors at $z\gtrsim 1$. Therefore, the mETG population at $z\sim 1$ is approximately equal to $\sim 135$\%$-95$\%$\sim 40$\% of the local mETG population, as the model predicts. This answers the question stated at the end of \S\ref{sec:comparison}. 

The interplay between wet, mixed, and dry mergers in the buildup of mETGs at $z\lesssim 1$ is shown in Fig.\,\ref{fig:etggrowthrate}. In the upper panel of this figure, we have plotted the predicted fractional buildup of the present-day number density of mETGs driven by the major mergers of each type at each redshift. The net fractional increment in the number density of mETGs at each redshift is marked with diamonds in the figure. The figure indicates that the major mergers reported by observations are enough to drive the appearance of $\sim 30$\% of the present-day number density of mETGs at $0.9<z<1$, an additional $\sim 20$\% at $0.8<z<0.9$, and finally, $\sim 10$\% at $0.7<z<0.8$. The previous fractions already consider the mETGs that are disappearing at each redshift due to dry mergers: numbers of mETGs equivalent to $\sim 15$\%, $\sim 10$\%, and $\sim 5$\% of the local mETGs population are being removed from the Universe through dry major mergers at $0.9<z<1$,  $0.8<z<0.9$, and $0.7<z<0.8$, respectively (circles in the figure).

In the lower panel of Fig.\,\ref{fig:etggrowthrate}, we have represented the accumulated numerical increment of the number density of mETGs driven by major mergers of each type, since each redshift down to the present. The net growth in the number density of mETGs driven by all major mergers is also plotted in the figure (diamonds). The figure poses that the interplay between the generation of "brand-new" mETGs through wet mergers, the removal of already-existing mETGs through dry mergers, and the conversion of existing mETG into more massive ones through dry and mixed mergers reproduces the  increment of the number density of mETGs observed since $z\sim 1$ ($\sim 60(^{70}_{50})$\% of their local value, see \S\ref{sec:numberevolution}).

\begin{figure}[!]
\begin{center}
\includegraphics*[width=0.5\textwidth,angle=0]{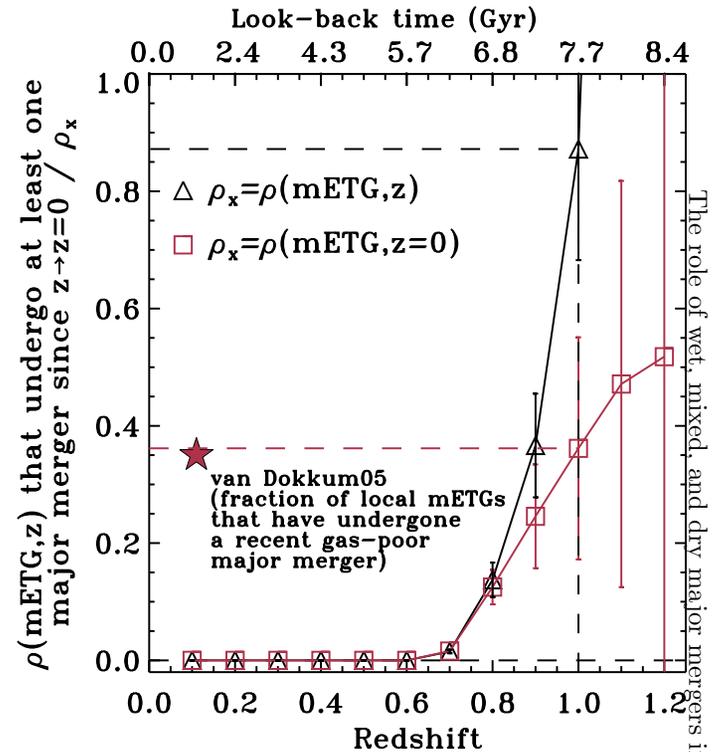}
\caption{Predicted fraction of mETGs that undergo \emph{at least} one gas-poor major merger (dry or mixed) between each redshift $z$ and the present, with respect to the local number of mETGs (\emph{squares}), and to the number of mETGs that exists at $z$ (\emph{triangles}). \emph{Star}: Fraction of local mETGs that have undergone a gas-poor major merger in their recent past, as derived by \citet{2005AJ....130.2647V}. This observational estimate coincides pretty well with the model prediction of the local number of mETGs that can have undergone a gas-poor major merger in the last $\sim 8$\,Gyr. \emph{[A color version of this plot is available at the electronic edition.]}
}\label{fig:ndrymixedperetgz}
\end{center}
\end{figure}

In Fig.\,\ref{fig:ndrymixedperetgz}, we represent the number of mETGs of each redshift that experience at least one gas-poor major merger (mixed or dry) from that redshift down to the present, according to the model. This number is provided with respect to the local number of mETGs (squares), and to the number of mETGs existing at that redshift (triangles). Notice that this quantity is a lower limit to the number of mETGs at each redshift that are going to take part as gas-poor progenitors in at least one gas-poor major merger since that redshift to the present. This means that at least $\sim 87(^{110}_{67})$\% of the mETGs existing at $z\sim 1$ are going to be involved as gas-poor progenitors in the major mergers occurred since $z\sim 1$ (triangles in the figure). Equivalently, we can derive that at least $\sim 35(^{55}_{15})$\% of local mETGs have acquired more than $\sim 1/3$ of its final mass through at least one gas-poor major merger in the last $\sim 8$\,Gyr (squares in the figure). Therefore, the model shows that $\lesssim 13$\% of the mETGs existing at $z\sim 1$ can have evolved without undergoing a major merger since then, or equivalently, that $\lesssim 5$\% of present-day mETGs can have evolved passively since $z\sim 1$ down to the present in reality.

\begin{figure}[!]
\begin{center}
\includegraphics*[width=0.5\textwidth,angle=0]{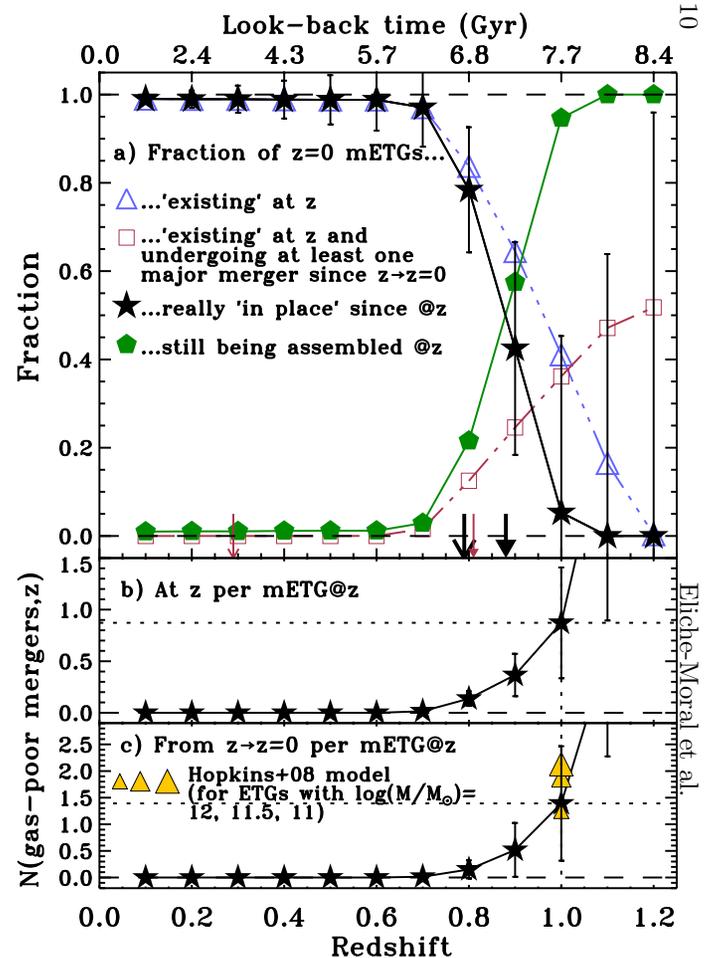}
\caption{
{\bf Panel a)\/}: Fraction of local mETGs in place since each redshift $z$ down to the present. \emph{Empty triangles}: Number density of mETGs predicted by the model at each redshift, referred to their present-day value (see Fig.\,\ref{fig:etgnumbergrowth}). \emph{Empty squares}: Fraction of mETGs at each redshift that undergo at least one gas-poor major merger (mixed or dry) since that redshift down to the present (see also Fig.\,\ref{fig:ndrymixedperetgz}). \emph{Filled stars}: Fraction of local mETGs that are really in place since each redshift to the present. \emph{Filled pentagons}: Fraction of present-day mETGs being assembled at each redshift. The errors of this distribution at each redshift are identical to those plotted on top of the distribution of filled stars. \emph{Thick arrows}: Redshifts of final assembly of 50\% and 80\% of present-day mETGs (filled and open arrows, respectively), according to our model. \emph{Thin arrows}: Redshifts at which 50\% and 80\% of the star content of present-day mETGs are enclosed into one single object according to \citet{2006MNRAS.366..499D} model (filled and open arrows, respectively). {\bf Panel b)\/}: Number of gas-poor major mergers undergone at each redshift per mETG existing at that redshift. {\bf Panel c)\/}: Average number of major mergers undergone by each mETG existing at each redshift since that redshift down to the present. \emph{Triangles}: Predictions by \citet{2008ApJS..175..390H} model on the number of gas-poor major mergers undergone per present-day ETG with present-day stellar masses $\log(\mathcal{M}_*/\Msun)=11$, 11.5, and 12. \emph{[A color version of this plot is available at the electronic edition.]} 
}\label{fig:fetginplace}
\end{center}
\end{figure}

\subsection{Present-day mETGs really in place since $z\sim 1$}
\label{sec:mETGinplace}

A galaxy "is in place" since a certain redshift basically if: 1) it has not received any significant external mass contribution since that redshift, 2) if its global structure has not changed dramatically since then, and 3) if its mass has increased negligibly through its SFH since that redshift. In the model context, a mETG is going to be in place since a given redshift basically if it does not experience any major merger since then down to the present, as ETGs exhibit so low SFRs at $z<1$ that we can consider the third condition fulfilled by them during the last $\sim 8$\,Gyr. 

We have seen that the model predicts that $\sim 87(^{110}_{67})$\% of the mETGs located at $z\sim 1$ are going to be the gas-poor progenitors in at least one major merger during the last $\sim 8$\,Gyr (see \S\ref{sec:gaspoormergers}). The remaining mETGs existing at $z\sim 1$ ($\sim 13$\%) do not experience major mergers since then, meaning that they must have been in place since then. Having into account that the number density of mETGs at $z\sim 1$ is $\sim 40$\% of the local one (\S\ref{sec:numberevolution}), the mETG population in place since $z\sim 1$ represents $\lesssim 5$\% of the local mETG population. Therefore, the fraction of present-day mETGs that have been in place since $z\sim 1$ is negligible, as derived from current observational merger fractions. Two questions arise then: first, what the redshifts of final assembly of present-day mETGs are, and secondly, how many present-day mETGs are really in place at each redshift since $z\sim 1$.

In the upper panel of Fig.\,\ref{fig:fetginplace}, we show the predicted fractions of local mETGs that are really in place since each redshift and that are still being assembled at each redshift, according to the model (filled stars and pentagons, respectively). As shown in \S\ref{sec:numberevolution}, the model reproduces the observational fact that the number density of mETGs at $z\sim 1$ is $\sim 40(^{50}_{30})$\% of the present-day one (empty triangles in Fig.\,\ref{fig:fetginplace}), but it also poses that the real number of local mETGs that have been in place since $z\sim 1$ is much lower, of $\lesssim 5$\% (filled stars in the same figure). The majority of mETGs located at $z\sim 1$ have played the role of gas-poor progenitor in any of the gas-poor major mergers occurred during the last $\sim 8$\,Gyr (compare the empty squares and the empty triangles at $z\sim 1$). Therefore, the model predicts that nearly all present-day mETGs are being assembled at $z\sim 1$, that more than a half of them are ongoing their assembly at $0.8<z<0.9$, and that $\sim 20$\% of them are still accreting mass through major mergers at $0.7<z<0.8$ (pentagons). At $z\sim 0.65$, nearly all mETGs are completely-assembled and in place in the Universe. In the last $\sim 5$\,Gyr since $z\sim 0.5$, the model predicts that the numerical growth of the mETGs through the major mergers reported by present observations is $<2$\% of its present value.

The model shows that the mETGs at $z\sim 1$ are not just the past, brighter counterparts of some present-day mETGs (passively-faded since then), but their gas-poor progenitors. Therefore, very few present-day mETGs have been really in place since $z\sim 1$ ($\lesssim 5$\%), considering the major merger fractions reported by current observations. 

The middle panel of the figure shows the number of gas-poor mergers undergone at each redshift per mETG located at that redshift. Nearly all mETGs existing at $z\sim 1$ are involved in a major merger only during the $\sim 0.4$\,Gyr time-period spanning at $0.9<z<1$. 

In the lower panel of Fig.\,\ref{fig:fetginplace}, we represent the average number of major mergers undergone since a certain redshift to the present per each mETG existing at that redshift. Each mETG existing at $z\sim 1$ undergoes an average of $\sim 1.4$ major mergers since $z\sim 1$, basically during the period elapsed at $0.7<z<1$, in excellent agreement with the predictions of the model by \citet{2008ApJS..175..390H} for $\log(\mathcal{M}_*/\Msun)=11$ galaxies. Referring this estimate to the number density of present-day $L>L^{*}$ galaxies, we would derive that each present-day bright galaxy has experienced $\lesssim 0.28$ gas-poor major mergers since $z\sim 1$. So, the model can explain the low observational probability of galaxies in general of having experienced more than one gas-poor merger \citep[][]{2006ApJ...644...54M}, as well as the high probability exhibited by mETGs of having experienced one dry merger since $z\sim 1$ \citep[see][]{2005AJ....130.2647V}.

\section{Discussion}
\label{sec:discussion}

\subsection{Reconciling the hierarchical scenario for the origin of mETGs with observations}
\label{sec:role}

The model demonstrates the feasibility of the mixed scenario proposed by \citet{2007ApJ...665..265F}, proving that the coordinated effects of wet, mixed, and dry major mergers since $z\sim 1$ are enough to explain the increase in the number density of mETGs (by a factor of $\sim 2.5$) observed since that epoch \citep{2009ApJ...697.1369B,2009MNRAS.395..144W}. Basically, we have seen that wet major mergers must have played the dominant role in this buildup \citep[in agreement with][]{2008ApJ...681..232L,2009A&A...498..379D,2009ApJ...702..307S,2009MNRAS.395..144W,2009arXiv0910.1598W}, although the contribution of dry and mixed mergers has been essential in order to reproduce it \citep{2009ApJ...697.1369B}. 

Moreover, the model shows that the merger fractions reported by observations imply directly that most of this recent assembly has occurred basically during the $\sim 1.4$\,Gyr time-period elapsed at $0.7<z<1$, as pointed by several observational studies \citep{2000MNRAS.311..565L,2009MNRAS.399L..16C,2010ApJ...709..644I}. Since $z\sim 0.7$, the contribution of major mergers in general (and of dry major mergers, in particular) to this assembly has been negligible. This explains why the number density of mETGs has remained nearly constant during the last $\sim 6.3$\,Gyr \citep[][]{2006ApJ...648..268B,2006A&A...453..809I,2006ApJ...644...54M,2007ApJ...666..212D,2007ApJS..172..494S,2007ApJS..173..432X,2009MNRAS.398.1549R,2010AJ....139..794D,2010arXiv1001.2015T}. 

However, other studies report a relevant role of major mergers in the buildup of mETGs at low redshift. \citet{2005AJ....130.2647V} derives that, if the ongoing mergers of local mETGs are representative for the progenitors of the remnants, $\sim 35$\% of mETGs at $z\sim 0.1$ have experienced a major merger in their recent past (usually with another gas-poor galaxy). We have overplotted his estimate in Fig.\,\ref{fig:ndrymixedperetgz}. It coincides pretty well with our predicted fraction of local mETGs that have experienced at least one gas-poor major merger since $z\sim 1$. So, the result by \citeauthor{2005AJ....130.2647V} could be compatible with ours if the tidal debris generated by the last gas-poor major merger experienced by a mETG remained "fossilized" during $\sim 4$-5 Gyrs around it. However, this seems not to be the case, as this timescale is estimated to be twice smaller \citep[$\sim 1$-2\,Gyr, see][]{2008ApJ...684.1062F,2009MNRAS.400.1264S}. It is also possible that both results can not be directly compared, as \citeauthor{2005AJ....130.2647V}'s sample is somewhat different to ours and his method is also sensitive to mergers of mass-ratios lower than $1:3$. Nevertheless, his estimate on the fraction of tidally disturbed local mETGs agrees with our predicted fraction of local mETGs appeared into the cosmic scenario since $z\sim 1$ (see Fig.\,\ref{fig:etggrowthrate}). 

Although the unreal high merger rates traditionally predicted by $\Lambda$CDM models have frequently been put forward against the hierarchical scenario, improvements in the definitions of the merging rates of halos and of the sub-baryonic clumps that reside within them are deriving theoretical major merger rates promisingly similar to observational estimates \citep{2006ApJ...652...56B,2007ApJ...666..212D,2007ApJS..172..494S,2008ApJ...688..789G,2008ApJS..175..356H,2008ApJS..175..390H,2008ApJ...683..597S,2009ApJ...702.1005S,2009MNRAS.396.2345B,2009ApJ...701.2002G,2009MNRAS.tmp.1955B,2010arXiv1001.2368B,2010arXiv1001.4533S}. We have represented in Figs.\,\ref{fig:nmergerstot1}, \ref{fig:nmergerstot2}, \ref{fig:nmergers}, and panel c) of Fig.\,\ref{fig:fetginplace} the predictions of several hierarchical models in the cold dark matter paradigm \citep{1996MNRAS.283.1361B,2006ApJ...640..241B,2008ApJ...680...41D,2008ApJS..175..390H,2009MNRAS.397..506K}. As shown in the figures, the hierarchical model predictions at different redshifts and for different merger types agree pretty well with our results, which assume observational major merger fractions. All these results indicate that the hierarchical framework for the origin of mETGs and observations may be surprisingly compatible afterwards. 

\subsection{Different mechanisms to generate disky/boxy ellipticals and lenticulars}
\label{sec:etgsaccordingtomergertype}

Explaining the difference between the formation of ellipticals (E's) and lenticular galaxies (S0-S0a's) is a new challenge for models of early-type galaxy formation \citep{2010MNRAS.tmp...24B}. \citet{2009arXiv0911.1126O} conclude that both galaxy types most likely have a different origin, as about 60-70\% of the observed transformations from late-type galaxies to ETGs must result in disky S0-S0a systems, being the remaining cases disk-less E's. Our model indicates that the mechanisms that have built up nearly all present-day mETGs (E's or S0-S0a's) since $z\sim 1$ have been the same (major mergers), but it also establishes naturally the possible mechanism that differences the generation of an E from that of a S0-S0a. This is the type of the last major merger that assembled the ETG in question.

Those mETGs deriving exclusively from a gas-rich merger probably had enough gas and angular momentum in their environments to rebuild a disk around the spheroidal remnant, ending as S0-S0a's \citep[see][and \S2.3.1 in \citetalias{2010arXiv1002.3537E}]{2005A&A...430..115H,2008ApJ...683..597S,2009arXiv0912.1077B}. However, if an already-formed ETG becomes the gas-poor progenitor of another major merger, the probability of rebuilding a disk in this case is much lower, because at least one of the progenitors is an already gas-depleted galaxy. So, the resulting ETG in dry or mixed mergers will probably resemble more an E than a S0-S0a. 

In \S\ref{sec:comparison2}, we have derived that all local mETG are susceptible of \emph{deriving} from one wet major merger since $z\sim 1$. However, only $\sim 35$\% of them have really \emph{taken part} in a major merger as gas-poor progenitors during the last $\sim 8$\,Gyr, after their first appearance as ETGs into the cosmic scenario (\S\ref{sec:gaspoormergers}). Therefore, only these $\sim 35$\% of present-day mETGs will exhibit finally an elliptical morphology, while the remaining local mETGs ($\sim 65$\%) derive exclusively from a wet major merger, and thus, they have preserved a disk until the present (ending up as S0-S0a's). These percentages reproduce the present-day relative fractions of E's and S0-S0a's indicated by \citeauthor{2009arXiv0911.1126O}, and support the results by \citet{2005AJ....130.2647V}, who concludes that the majority of today's most luminous field ellipticals have been recently built up through gas-poor, major mergers.

This scenario also provides a feasible explanation to two recent results derived from the physical properties of local bright ellipticals: 1) the fact that these properties point preferentially to a last gas-poor major merger \citep{2003ApJ...597L.117K,2009arXiv0911.0044R}, and 2) to the fact that their central physical characteristics favour a formation scenario in which both dissipational and dissipationless processes must have played a role \citep[as occurs in our scenario, see][]{2010arXiv1002.0585L}. It can also explain the presence of small amounts of young stars and of nuclear dust in ETGs  \citep{1995AJ....110.2027V,2000ApJ...541L..37F,2007ApJS..173..619K}, as well as the fact that SF-downsizing appears to be accelerated in overdense regions hosting higher numbers of massive ETGs, as  compared to the underdense regions \citep[][]{2006ApJ...651..120B,2006MNRAS.370..828F}. Evidence of the preferential transformation of spirals into S0's through major mergers instead of through galaxy-environment processes has been also reported in clusters at $0.1<z<0.8$ \citep{2010ApJ...711..192J}, in agreement with our scenario. Moreover, this formation mechanism for E's and S0-S0a's is also supported by simulations, that show that a major spiral-spiral remnant exhibits morphological and kinematical properties which resemble those of intermediate-mass S0-S0a's \citep[][]{2003ApJ...597..893N,2006ApJ...641...90R,2007MNRAS.379..401E}, whereas the E's fit better as remnants of dry major mergers \citep{2006ApJ...636L..81N}. 

We have seen that there have been $\sim 0.95$ gas-poor major mergers per present-day mETG since $z\sim 1$, of which $\sim 1/3$ correspond to dry events, being the remaining $\sim 2/3$ associated to mixed events (\S\ref{sec:comparison2}). These fractions are quite similar to the relative percentages of slow-rotating, boxy ellipticals (with cusped-central profiles) and of rapid-rotating, disky ellipticals (with cored-central profiles) found at the present with $11<\log(\mathcal{M}_*/\Msun)<11.5$ \citep[see][]{1992ApJ...399..462B,1993ApJ...411..153B,2007ApJ...664..226L,2007ApJ...664..738P}. So, it is possible that the dichotomy observed in the properties of $z=0$ ellipticals is determined by the gas-content of the last gas-poor major merger that assembles a present-day elliptical. If it is a mixed event, the final mETG will be a rapid-rotating, disky elliptical, whereas if the event is dry, the final mETG will be a slow-rotating, boxy elliptical. This scenario is quite similar to that proposed by \citet{2008ApJS..175..390H}.

Notice that the model can also explain why basically all present-day quiescent galaxies with $\mathcal{M}_* \gtrsim 10^ {11}\Msun$ are essentially spheroidal \citep{2009ApJ...706L.120V}, as it predicts that nearly all local mETGs have been built up through at least one (wet) major merger in the last $\sim 8$\,Gyr, a process that inevitably results in big spheroids.

Obviously, the model does not exclude the contribution of other mechanisms to the formation of S0-S0a's, such as the effects of the intracluster medium on the infallen star-forming galaxies or the minor mergers \citep[see][]{2006A&A...458..101A,2006A&A...457...91E,2008A&A...483L..39C,2009ApJ...693..112P,2009MNRAS.395L..62G,2009MNRAS.394.1713K,2009ApJ...695....1T,2009MNRAS.393.1302W,2010arXiv1002.2145L}. However, observations indicate that their role has been more significant in galaxies of lower masses \citep[\ie, for $\log(\mathcal{M}_*/\Msun)<11$ systems, see][]{2004ApJ...603L..73X,2009ApJ...697.1369B,2009arXiv0901.4545D}. 

\subsection{Redshift of complete assembly of mETGs}
\label{sec:zassembly}

As commented in \S\ref{sec:introduction}, hierarchical models predict that mETGs have been the latest systems to complete their assembly in the Universe. Therefore, the knowledge of the redshift at which these systems are "in place" in the cosmic scenario is essential to refute or support the hierarchical paradigm of galaxy formation \citep[see, \eg,][]{2006MNRAS.366..609C,2007ApJS..172..494S,2007ApJS..172..406S,2009arXiv0911.1126O,2010ApJ...709..644I,2009arXiv0907.5416P}. 

In this sense, the observational phenomenon of galaxy mass-downsizing \citep[][]{2004Natur.430..181G,2005ApJ...630...82P,2008ApJ...675..234P,2008A&A...491..713W} has been usually interpreted incorrectly as an obstacle to hierarchical theories. First, because the model is capable of reproducing the observed evolution of the massive-end of the galaxy LFs up to $z\sim 1$ just accounting for the effects of the observed major mergers, in general agreement with mass-downsizing trends \citepalias[as shown in][]{2010arXiv1002.3537E}. And secondly,
because the model shows that the classical interpretation consisting on that the mETGs existing at a given redshift "have been in place since then" is completely wrong, as most of the mETGs existing at $z\sim 1$ are predicted to have been the gas-poor progenitors of the mixed and dry major mergers that have assembled present-day mETGs during the last 8\,Gyr (\S\S\ref{sec:gaspoormergers}-\ref{sec:mETGinplace}). 

The model predicts that the present-day population of mETGs has not been finished their assembly at least until $z\sim 0.6$ (see panel a of Fig.\,\ref{fig:fetginplace}). This result agrees with the evolution of the galaxy mass functions reported by \citet{2008ApJ...675..234P}. Accounting for the time it takes for a post-merger galaxy to migrate from the blue cloud to the red sequence  \citep[$\sim 0.5$-1\,Gyr, see][]{1999ApJ...517..130H,2006A&A...454..125E,2004ApJ...607L..87C,2007MNRAS.382..960K,2008MNRAS.384..386C,2008MNRAS.391.1137L}, the previous assembly redshift moves to lower redshifts. Accounting for this, the model predicts that the whole present population of mETGs must be completely relaxed at $z\sim 0.5$, a result that is in excellent agreement with the predictions of several hierarchical models  \citep{2007MNRAS.375....2D,2009arXiv0911.1126O} and with the assembly redshifts derived from studies of the stellar populations of the brightest cluster ellipticals \citep{2000ESASP.445...37L,2008MNRAS.388.1537M,2008ApJ...683L..17T}. 

It is worth noting that, making a "traditional" interpretation of the model results, we would derive that present-day mETGs have already assembled half of their stellar masses at $z\sim 1$ (see triangles in Fig.\,\ref{fig:fetginplace}), in agreement with observations supporting mass-downsizing \citep[see Fig.\,6 by][]{2008ApJ...675..234P}. However, attending to when half of the present-day population of mETGs are really in place (\ie, not experiencing any other major merger down to $z=0$), we would derive that this epoch occurs $\sim 0.7$\,Gyr later, at $z\sim 0.85$ (see stars in Fig.\,\ref{fig:fetginplace}). Considering an additional post-merger period for galaxy relaxation of $\sim 0.5$-1\,Gyr as commented above, the assembly redshift for half of present-day mETGs derived by our model is predicted to occur $\sim 2$\,Gyr later than the estimate derived through a "traditional" interpretation of observations (it moves from $z\sim 1$ to $z\sim 0.65$). 

In Fig.\,\ref{fig:fetginplace}, we have marked the redshift at which the model by \citet{2007MNRAS.375....2D} predict that 50\% of the stars of local ETGs are enclosed into one single body ($z\sim 0.8$). Accounting for the offset of $\sim 0.5$-1\,Gyr for system relaxation, the redshift at which our model predicts that $\sim 50$\% of mETGs are in place in the Universe coincides pretty well with the prediction by \citeauthor{2007MNRAS.375....2D}, even if we consider that both quantities are not exactly equivalent (compare the corresponding symbols in the figure). However, notice that these author predict that $\sim 20$\% of the stellar mass of local mETGs is still to be assembled at $z\sim 0.3$, whereas our model predict that $>98$\% of local mETGs have already in place since $z\sim 0.45$-0.5, accounting for the previously-commented offset. The extremely late epoch of mETG assembly reported by \citeauthor{2007MNRAS.375....2D} has its observational counterpart, as the massive end of the galaxy LF is observed to be still evolving through major mergers (basically, dry) at a measurable level at $z < 0.7$ in high-density regions \citep{2008MNRAS.388.1537M,2009MNRAS.396.2003L}. However, our model reproduces the galaxy evolution of field mETGs (see Paper I). As the mass assembly evolution and the merger rate are known to depend strongly on the environment 
\citep{1995AJ....110..129K,2008ApJ...683L..17T,2009ApJ...700..791H,2010MNRAS.402..447R,2009A&A...503..379T,2009A&A...508.1217Z,2009arXiv0910.0245C,2009ApJ...701..994W,2010arXiv1001.4560L,2010arXiv1002.0835R}, it is possible that our model is not sensitive to the extremely late phase of mETGs buildup occurring in clusters \citepalias[see][]{2010arXiv1002.3537E}. Note that the model predictions do not rule out either a possible early period of rapid growth at $z>1.5$ for some brightest cluster ellipticals \citep[rather than a prolonged hierarchical assembly since $z\sim 1$, see][]{2007MNRAS.377.1717K,2008ApJ...680...41D,2009Natur.458..603C,2009ApJ...691L..33K}, as these systems (with $L>3L^*$) represent less than 5\% of our $L>L^*$ sample. 

In conclusion, the model poses a revealing, new insight into the question of the final assembly redshift of present-day mETGs: the consideration of that the mETGs at $z\sim 1$ may not be the passively-evolved, high-z counterparts of some present-day mETGs, but their gas-poor progenitors instead. This moves down the redshift of complete assembly of this galaxy population to $z\sim 0.5$, in agreement with predictions of hierarchical models, meaning that a correct interpretation of observations is a key point to reconcile the late buildup of mETGs predicted by hierarchical models with the observational mass-downsizing phenomenon \citep[see also][]{2009MNRAS.397.1776F}.

\subsection{Mass growth in the red sequence driven by major mergers}
\label{sec:redsequence}

The rise of the number density of mETGs since $z\sim 1$ by a factor of $\sim 2$ predicted by the model entails that the stellar mass of the red sequence has nearly doubled during the last $\sim 8$\,Gyr (\S\ref{sec:numberevolution}). This prediction is in agreement with independent observational estimates \citep{2004ApJ...608..752B,2005AJ....130.2647V} and with predictions of semianalytical models in a CDM scenario \citep{2007MNRAS.375....2D}. Moreover, accounting for the fact that ETGs represents nearly $\sim 50$\% of the local massive galaxy sample \citep{2010MNRAS.tmp...24B}, the previous prediction implies that the mass accreted in major mergers since $z\sim 1$ amounts to $\sim 25$\% of the total stellar mass density of the massive systems at $z=0$. This value is also similar to those derived from observational estimates \citep{2008ApJ...687...50P,2009ApJ...697.1369B,2009A&A...498..379D}, although a bit higher than the value of $\sim 15$\% reported by \citet{2000ApJ...532L...1C}.

The model predicts that there have been $\gtrsim 0.35$ dry major mergers per local mETGs since $z\sim 1$ (see \S\ref{sec:comparison2}). This means that dry major mergers are responsible of the accretion of $\sim 35$\% of the present total stellar mass of mETGs, in agreement with the value of $\sim 38$\% reported by \citet{2010arXiv1001.4560L}. This is equivalent to say that gas-poor mergers occurred at $z\lesssim 1$ are responsible of accreting $\sim 18$\% of the total stellar mass density of local massive galaxies, an estimate that is coherent with that obtained by \citet{2009MNRAS.397..506K} using a CDM-based semianalytical model ($\sim 10$-20\%). Moreover, as our model predicts that there have been $\sim 0.85$ gas-poor major merger (dry or mixed) per local mETGs during the last $\sim 8$\,Gyr (\S\ref{sec:comparison2}), local mETGs can not have assembled more than $\sim 45$\% of their stellar mass via dissipationless mergers, in excellent agreement with the prediction derived by \citet{2009ApJ...706L..86N} too.

All these results stress the feasibility of reproducing the bulk of the observed mass transfer from the massive-end of the blue galaxy cloud to that of the red sequence since $z\sim 1$, just accounting for the merger fractions reported by current observations. Secular processes probably also play a role in this process, although their effects must be insignificant as compared to those of major mergers in the high-mass range. In fact, observations indicate that secular processes seem to have contributed more relevantly to the formation of intermediate-mass Sa-Sb disks than to the buildup of mETGs \citep{2008mgng.conf...47A,2008ASPC..396..325C,2008ASPC..396..297K,2009ApJ...697.1971J,2009arXiv0912.1077B,2010ApJ...710.1170L,2009A&A...505..497Y}. Therefore, the model not only supports the key role played by major mergers in the assembly of mETGs since $z\sim 1$ \citep{2005ApJ...625..621B,2007A&A...476..137A}, but it also shows that hierarchical model predictions and observations agree pretty well in this sense.

\section{Summary and conclusions}
\label{sec:conclusions}

In this paper, we analyse the relative role of wet, mixed, and dry major mergers in the recent buildup of mETGs, as derived from the model presented in \citetalias{2010arXiv1002.3537E}. The model traces back-in-time the evolution of the local galaxy populations considering the number evolution derived from observational merger fractions and the L-evolution of each galaxy type due to typical SFHs. Several relevant results concerning to the recent buildup of mETGs have been derived:\\[-0.2cm]

\indent 1. The model demonstrates the feasibility of the mixed scenario proposed by \citet{2007ApJ...665..265F}, proving that the coordinated effects of the wet, mixed, and dry major mergers strictly reported by current observations since $z\sim 1$ can explain the increase by a factor of $\sim 2.5$ observed in the number density of mETGs since then. \\[-0.3cm]

\indent 2. There have been $\sim 2$ major mergers per local mETG since $z\sim 1$, $\sim 1$ corresponding to wet mergers and the another one to dry$+$mixed mergers. Therefore, although wet major mergers have played the dominant role in the recent buildup of mETGs, the contribution of dry and mixed mergers has also been essential in it.\\[-0.3cm]

\indent 3. The bulk of this mETGs assembly is predicted to take place during a $\sim 1.4$\,Gyr time-span at $0.7<z<1$. Since $z\sim 0.7$, the number density of mETGs has remained nearly constant. The model shows that this frostbite in the assembly of mETGs is basically driven by the decrease of the major mergers fraction registered during the last $\sim 6.3$\,Gyr. The effect of a higher relative fraction of dry and mixed mergers than of wet mergers at $z<0.8$ on this frostbite is secondary and irrelevant.\\[-0.3cm]

\indent 4. The stellar mass of the red sequence has nearly doubled through major mergers during the last $\sim 8$\,Gyr. Dry major mergers since $z\sim 1$ can account for the assembly of $\sim 35$\% of the total stellar mass of local mETGs. These estimates are in excellent agreement with those derived from observations and from hierarchical models, stressing the feasibility of reproducing the observed mass transfer from the massive end of the blue galaxy cloud to that of the red sequence since $z\sim 1$, just accounting for the major mergers reported by current observations. \\[-0.3cm]

\indent 5. The model poses a revealing, new insight into the question of the final assembly redshift of present-day mETGs: at least $\sim 87(^{110}_{67})$\% of the mETGs existing at $z\sim 1$ are not the passively-evolved, high-z counterparts of the corresponding number of present-day mETGs, but their gas-poor progenitors instead. This implies that $\gtrsim 95$\% of local mETGs have acquired more than $1/3$ of their present masses through major mergers during the last $\sim 8$\,Gyr. Therefore, $\lesssim 5$\% of present-day mETGs have been really in place since $z\sim 1$.\\[-0.3cm]

\indent 6. Nearly all present-day mETGs are being assembled through major mergers at $0.9<z<1$, $\sim 50$\% of them are ongoing their assembly at $0.8<z<0.9$, and $\sim 20$\% of them are still accreting mass through major mergers at $0.7<z<0.8$. At $z\sim 0.6$, the present population of mETGs is completely-assembled in the Universe. Accounting for the time it takes for a post-merger galaxy to migrate from the blue cloud to the red sequence ($\sim 0.5$-1\,Gyr), present-day mETGs must be completely in place at $z\sim 0.5$, in agreement with the predictions of hierarchical models and with studies of the stellar populations of mETGs. So, accounting for the fact that a mETG existing at a certain redshift is not necessarily in place since then, we can reconcile the late buildup of mETGs predicted by hierarchical models with observations supporting galaxy mass-downsizing. \\[-0.3cm]

\indent 7. The model predicts the present-day fractions of E's and S0-S0a's if ellipticals derive from a wet major merger plus at least one gas-poor event (dry or mixed) since $z\sim 1$, while local S0-S0a galaxies have been assembled just through one wet major merger during the same epoch. It can also reproduce the local fractions of rapid-rotating, disky and slow-rotating, boxy E's if the former ones derive from a last mixed major merger and the last ones from a last dry event. \\

Some specific model predictions are the following:\\[-0.2cm]

\indent i. The population of mETGs at $z\sim 1$ was being doubled at that epoch just through wet major mergers occurring at that epoch. Nearly $85$\% of mETGs at $z\sim 1$ were taking part as gas-poor progenitors in a major merger at $0.9<z<1$. \\[-0.3cm]

\indent ii. The number of major mergers taking place at $z=1$ is equivalent to $\sim 35(^{55}_{20})$\% of the present-day number of $L\gtrsim L^*$ galaxies ($\sim 5$\% correspond to dry mergers, $\sim 10$\% to mixed events, and $\sim 20$\% to wet mergers). This means that $\sim 0.85(^{1.50}_{0.50})$ major mergers were taking place at $z\sim 1$ per local mETG ($\sim 0.55$ corresponding to wet events, $\sim 0.23$ to mixed, and $\sim 0.07$ to dry). \\[-0.3cm]

\indent iii. The predicted number of major mergers occurred since $z\sim 1$ per local $L>L^*$ galaxy amounts to $\sim 65(^{110}_{40})$\%. According to the model, there have been $\sim 0.35$ wet major mergers, $\sim 0.2$ mixed, and only $\sim 0.1$ dry events per local $L>L^*$ galaxy. \\[-0.3cm]

\indent iv. There have been $\sim 1.8(^{2.6}_{1.2})$ major mergers since $z\sim 1$ per local mETG. Approximately one of these $\sim 2$ mergers takes place during a period of $\sim 0.5$\,Gyr elapsed at $0.9<z<1$, while the other one basically occurs during the period of $\sim 1$\,Gyr elapsed at $0.7<z<0.9$. Of these $\sim 2$ mergers occurred since $z\sim 1$, $\sim 0.35$ correspond to dry and $\sim 0.6$ to mixed events. \\[-0.3cm]

\indent v. The model indicates that at least $\sim 35(^{55}_{15})$\% of local mETGs must have undergone a gas-poor major merger in the last $\sim 8$\,Gyr. Equivalently, $\sim 87(^{110}_{67})$\% of the mETGs existing at $z\sim 1$ have taken part as gas-poor progenitors at least in one major merger since $z\sim 1$. \\[-0.3cm]

\indent vi. The major mergers reported by present observations are enough to drive the numerical appearance into the cosmic scenario of $\sim 30$\% of the present-day number density of mETGs at $0.9<z<1$, an additional $\sim 20$\% at $0.8<z<0.9$, and finally, $\sim 10$\% at $0.7<z<0.8$. These fractions already consider the mETGs that are disappearing at each redshift due to dry mergers: numbers equivalent to $\sim 15$\%, $\sim 10$\%, and $\sim 5$\% of the local mETGs population are being removed from the Universe through dry major mergers at $0.9<z<1$,  $0.8<z<0.9$, and $0.7<z<0.8$, respectively.\\

The model not only supports the key role played by major mergers in the assembly of mETGs since $z\sim 1$, but also shows that the hierarchical framework for the buildup of mETGs and observations at $z\lesssim 1$ are surprisingly compatible afterwards. A correct interpretation of observations is essential to reconcile the hierarchical assembly of mETGs with galaxy mass-downsizing. 

\begin{acknowledgements}
We wish to acknowledge J.\,P.Gardner for making its \ncmod\/ code publicly available. We also thank Antonio Ben\'{\i}tez, Guillermo Barro, Carlos L\'{o}pez-Sanjuan, Marc Balcells, Rafael Guzm\'{a}n, and Juan Carlos Mu\~{n}oz-Mateos for interesting discussions and comments. Supported by the Spanish Programa Nacional de Astronom\'{\i}a y Astrof\'{\i}sica, under projects AYA2006-02358, AYA2009-10368, and AYA2006–12955. MCEM acknowledges support from the Madrid Regional Government through the ASTRID Project (S0505/ESP-0361), for development and exploitation of astronomical instrumentation (http://www.astrid-cm.org/). Partially funded by the Spanish MICINN under the Consolider-Ingenio 2010 Program grant CSD2006-00070: "First Science with the GTC" (http://www.iac.es/consolider-ingenio-gtc/). This work is based in part on services provided by the GAVO data center.
\end{acknowledgements}

\bibliography{elic0709_def.bib}{}
\bibliographystyle{apj}{}

\end{document}